 \documentclass[preprint]{aastex}

\slugcomment{To be submitted to the Astronomical Journal.}

\shorttitle{NGC 4650A: a disk in formation}
\shortauthors{Iodice et al.}

\begin{document}


\title{The puzzle of the polar structure in NGC~4650A}


\author{E. Iodice\altaffilmark{1}}
\affil{International School for Advanced Studies, via Beirut 2-4, 34014
Trieste, Italy}

\author{M. Arnaboldi\altaffilmark{2}, G. De Lucia\altaffilmark{2}}
\affil{Osservatorio Astronomico di Capodimonte, via Moiariello 16, 80131
Napoli, Italy}

\author{J.S. Gallagher, III\altaffilmark{4} and L. S. Sparke\altaffilmark{4}}
\affil{University of Wisconsin, Department of Astronomy, 
475 N. Charter St., Madison, WI 53706-1582, U.S.A.}

\author{K.C. Freeman\altaffilmark{3}}
\affil{RSAA, Mt. Stromlo, Canberra, Cotter Road, Weston ACT 2611, Australia}




\begin{abstract}

This work presents new surface photometry and two-dimensional 
modeling of the light distribution of the Polar Ring Galaxy NGC~4650A, 
based on near-infrared (NIR) observations and high resolution 
optical imaging acquired during the Hubble 
Heritage\footnote{Observations obtained 
with the NASA/ESA {\it Hubble Space Telescope},
obtained at the Space Telescope Science Institute, which is operated by 
the Association of Universities for Research in Astronomy (AURA), Inc. 
under NASA contract NAS 26555.} program.

The NIR and optical integrated colors of the S0 and the polar ring,
and their scale parameters, are compared with those for standard galaxy
morphological types. The polar structure appears to 
be a disk of a very young age, while the colors and light 
distribution of the host galaxy do not resemble that of a 
typical early-type system.
We compare these observational results with the predictions from
different formation scenarios for polar ring galaxies.  The
peculiarities of the central S0 galaxy, the polar disk structure and
stellar population ages suggest that the polar ring galaxy NGC~4650A
may be the result of a dissipative merger event, rather than of an
accretion process.           
\end{abstract}                                                                


\section{Introduction}\label{intro}
Polar ring galaxies (PRGs) are early-type galaxies
surrounded by an outer ring, made up by gas, stars and dust, which
orbits in a nearly perpendicular plane to the equatorial one of the
central galaxy (Whitmore et al. 1990).
NGC~4650A is considered the prototype of the class of wide polar ring
galaxies, because the two components are bright and well-defined
(Whitmore 1991). 
This object is one of the best-investigated polar ring galaxy, 
in particular to put limits on the 3-D shape of its dark matter 
halo (Whitmore et al., 1987; Sackett \& Sparke, 1990; 
Sackett et al., 1994 and Combes \& Arnaboldi, 1996). 
To this aim, new high-resolution 21-cm observations
were carried out by Arnaboldi et al. in 1997 at the Australia 
Telescope Compact Array. These were meant to resolve the rotation
curve at small radii, where velocities were measured
into two planes, the polar ring and the host galaxy equatorial plane, 
and to reach out at larger distances into the halo.
The results showed that the total HI mass in this system is 
$8 \times 10^{9}$ $M_{\odot}$, and the HI distribution and kinematics 
is consistent with the one of a spiral disk.
The spiral pattern proposed by Arnaboldi et al. (1997) seems to be
confirmed by the recent HST image of NGC~4650A, acquired during the
Heritage project (Gallagher et al. 2001).

Is a polar disk likely to form in the standard picture
for the formation and evolution of polar ring galaxies?
The standard scenario suggests that a 
gas-rich dwarf galaxy was either accreted
by an early-type galaxy, or gas was stripped from a nearby
gas-rich object during a high speed encounter.  
The accreted material would then form a ring
(Quinn 1991, Hernquist \& Weil 1993), which settles into
one of the principal plane of the gravitational potential associated
with the host galaxy (Heisler, Merritt and Schwarzschild, 1982; 
Bertola et al. 1991).
Several studies focused on the evolution and stability of
highly inclined rings.  Smoothed-particle hydrodynamic simulations 
with strong dissipative cooling by Katz \& Rix (1992) and 
Christodoulou et al. (1992) 
were used to study the evolution of low mass, highly inclined ring.
Katz \& Rix (1992) and Christodoulou et al. (1992) showed
that a narrow ring precesses very slowly in a
quasi-equilibrium configuration, with no appreciable evolution over an
Hubble time.
The gas stripping scenario was studied by Reshetnikov \& Sotnikova (1997).
They analyzed the different morphologies generated
in high speed encounters between an elliptical or an S0 galaxy, and a gas
rich disk.  Their results indicate that the average radius
at which the ring forms is related to the central mass (luminous +
dark) concentration of the host galaxy.  If the mass is highly
concentrated, the ring forms at smaller radii; if the host galaxy
has an extended massive halo, the ring average radius ($\bar{R}$) can
be as large as 30~kpc.
The common characteristics of all these secondary events, 
i.e. accretion of a dwarf or gas stripping, are the formation 
of narrow polar annuli, whose radial extent is of the order of 10\% 
of the average radius, for a quasi-stable configuration, 
and the total amount of accreted gas is up to $10^{9}$ $M_{\odot}$. 

A quite different approach to the formation of polar ring galaxies was
proposed recently by Bekki (1998). In this scenario the polar ring
results from a ``polar'' merger of two disk galaxies with 
unequal mass.  
The ``intruder'', on a polar orbit with respect to the ``victim'' disk, 
passes through it near its center: it is slowed down, and pulled back 
toward the victim, by strong dissipation which is caused by the interaction 
with the victim gaseous disk.  
In this encounter the two galaxies must have a small relative
velocity, so that the intruder is brought to rest at the center of the
victim's disk.  
Dissipation removes the random kinetic energy of the
gaseous component of the victim's disk, so that some gas
can settle again into a disky configuration.
The morphology of the merger remnants depends on the merging
initial orbital parameter and the initial mass ratio of
the two galaxies. 
Bekki's scenario successfully reproduces many of the observed morphologies 
for polar ring galaxies, such as the existence of both 
wide and narrow rings, helical rings and double rings (Whitmore 1991).
When wide polar rings are produced they have the following characteristics: 
1) there is no central hole in the  polar structure;
2) mass distributions in the central component and ring
become more centrally concentrated after the encounter;
an R$^{1/4}$ profile can develop if the ``intruder'' disk is much more
massive than the ``victim'';
3) the central component is nearly gas-free, similar to an S$0$-like system,
while the density wave triggered by the intruder into the victim
disk causes rapid star formation, within $\simeq 10^9$ yr,
so that the polar structure is characterized by a very young stellar 
population.
This kind of model (Sec. 3.2.5 in Bekki's 1998 paper) does 
predict peculiar characteristics for the intruder too: 
the intruder experiences both a heating of the disk (it puffs up) and 
energy dissipation, 
so that its density should increase relative to standard unperturbed
stellar disks. 
While the `gas-accretion' scenario can account for the existence of
rings, the Bekki scenario would also explain the presence of wide and 
massive polar disks.  
One uncertainty related to the Bekki scenario is whether the
polar rings formed in this way are stable, i.e. for how long these
objects can preserve the polar morphology after the merger
remnant reaches virial equilibrium. 

To perform a detailed test of these two scenarios for the formation of
PRGs we need to have data with the highest angular resolution,
to resolve the inner central morphology of the polar ring 
and the central host galaxy, plus the NIR data, to probe the stellar population 
without the strong absorption caused by dust.
Such a data set is now available for NGC~4650A and the aim of this work is to
compare the predictions from different formation scenarios with
the observational results from a study of the new NIR and
high resolution HST data.
We adopt a distance of 41 Mpc based on 
$H_{0}= 70$ km $\mbox{s}^{-1} \mbox{Mpc}^{-1}$ and an heliocentric radial 
velocity $V = 2861$ km $\mbox{s}^{-1}$.  
 
The observational data set in the NIR and optical bands are described 
in Section~\ref{obs}. The morphology of the 
host galaxy and polar structure are discussed in Section~\ref{morphology}. 
The photometry and scale parameters for this system are derived in
Section~\ref{photo}. 
In Section~\ref{stellar} the integrated colors of the
polar structure and S0 are compared with those predicted from stellar
synthesis population models.
The new observational evidence coming from these data are summarized
in Section~\ref{formation}, and conclusions are derived.

\section{Observations}\label{obs}
Here the near-IR (J, H and Kn bands) and the
optical (B, V and I bands) observations obtained for the Polar Ring 
Galaxy NGC~4650A are presented.
The morphology of the two main components, the host galaxy and the ring,
are derived and discussed.

\subsection{NIR observations}\label{NIRobs}
\begin{figure}
\epsscale{0.6}
\plotone{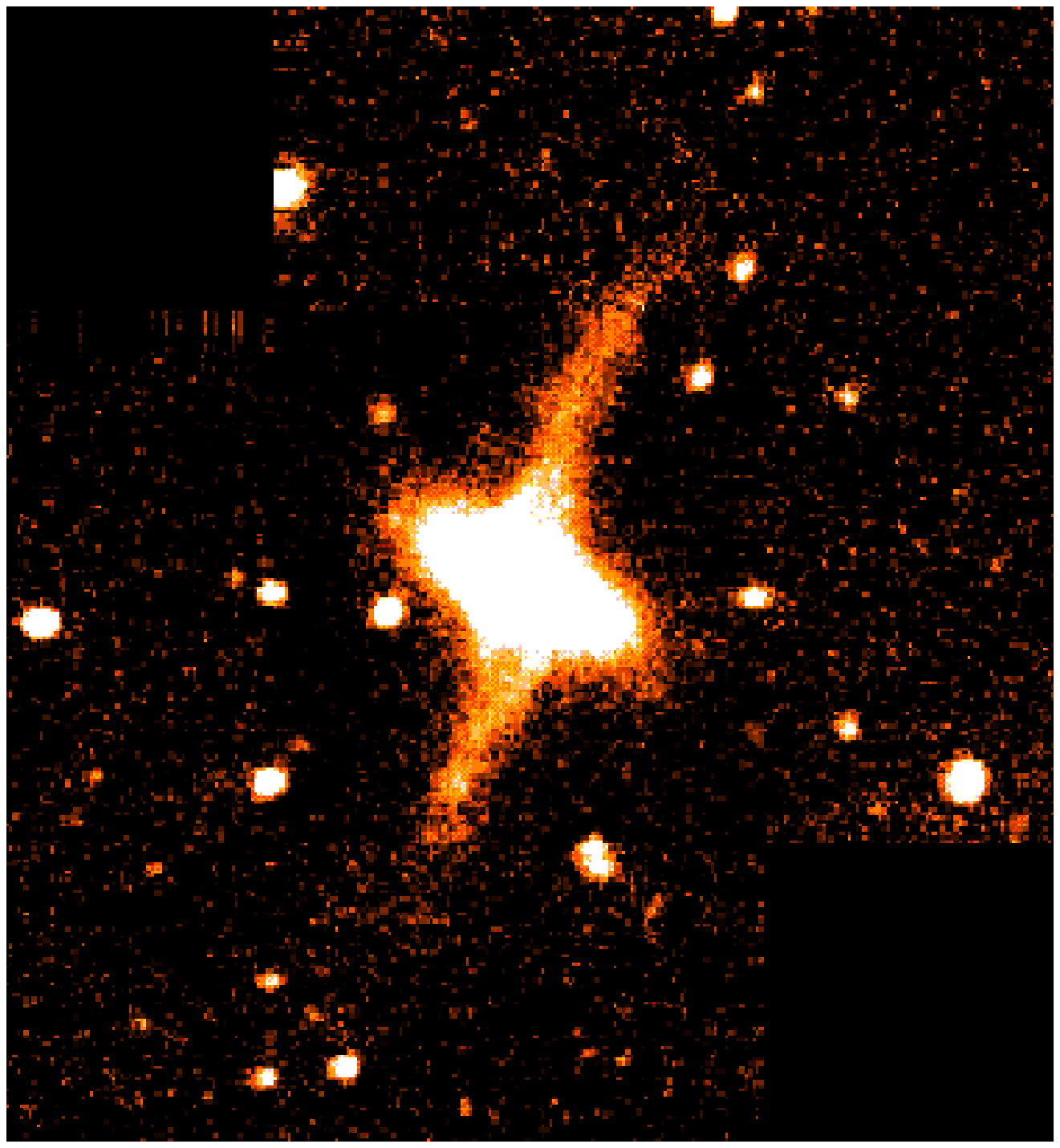}
\caption{\label{fig1} NGC~4650A in the Kn band, North is up and East is to
the left. The image size is $3'.1 \times 3'.0$ .}
\end{figure}

The near-infrared J, H and Kn images for NGC~4650A were obtained during two
observing runs at the Mt. Stromlo and Siding Spring Observatory 2.3 m
telescope with the CASPIR infrared camera (McGregor, 1994), 
with a field of view
of $2'.0 \times 2'.0$ and an angular resolution of 0.5 arcsec per pixel.
The observing log for these data is given in Table~\ref{tab1}.
Images were acquired with the offsetting mode and a cycle was defined 
containing 5 images on target plus 5 sky frames. 
Four cycles were obtained for J and H bands, and 8 cycles for Kn band were 
needed to have a better estimate of the background level.
Linearisation, flatfielding, sky-subtraction and bad pixel correction were
performed using the REDIMAGE task in the CASPIR package in IRAF. The resulting 
image for each cycle was derived by registering and combining all sub-frames. 
The final image  in each band was obtained by stacking images from each 
set of cycles.
Several standard stars, from Carter \& Meadows (1995), were observed at the
beginning, middle and at the end of each night, in order to transform
the magnitudes into the standard J, H and Kn band systems. 
The zero points that we derived are in good agreement with the indicative values
derived for the CASPIR camera (available in the CASPIR user manual) in each 
band; the differences in zero points were less than 0.1 mag.
The ring in NGC~4650A is too extended to lie inside one single pointing,
therefore it was necessary to take two different frames and mosaic them in
a single image. 
The two pointings are reduced independently; before the final
mosaic image was produced we checked that the two pointings have similar values
of the background level. 
Fig.~\ref{fig1} shows the final Kn band mosaiced image. 
For the combined images in all bands, several tests were performed on the 
background noise. 
Compared with the pixel-to-pixel variation, we 
found that there is more noise, both within sky regions of moderate
size and between such regions, than predicted by a pure counting
(Poisson) model.  
Large scale variations in the background, due to an imperfect flatfield correction,
contributed significantly to this ``extra noise''.
For the areas within which we measured magnitudes and colors (see
Sec.\ref{photo} below), we derived an estimate of the ``total'' error 
of the flux enclosed within the polygons, 
from the statistics of the sky background in many sky boxes of comparable area.
The {\it standard deviation for each pixel} is the standard deviation 
of the mean counts per pixel in each measured areas.

\subsection{Hubble Space Telescope observations}\label{HSTobs}
NGC~4650A was observed with WFPC2 on the Hubble Space Telescope (HST) 
in April 1999, during the Hubble Heritage Project 
(Gallagher et al. 2001).
The filters used are F450W, F606W and F814W, at 
optical wavelengths.
Fig.~\ref{fig2} shows the final multicolor image of NGC~4650A from the 
HST press release.
The observing log is listed in Table~\ref{tab2}.
Several frames were taken for each filter with different exposure times.
During data reduction, different exposures were registered,
scaled to the same exposure time and combined to obtain the final image frame
after cosmic rays removal. 
The magnitudes for the F450W, F606W, F814W filters were computed
following Holtzman et al. (1995b). 
For each filter $i_{\lambda}$, these are given by
\begin{equation}
m(i_{\lambda})=-2.5log(DN/s)+2.5log(GR)+ZP[m(i_{\lambda})]
\end{equation}
where GR is the gain ratio,
DN is the data number per second and $ZP[m(i_{\lambda})]$
is the filter zero point, from Table~9 of Holtzman et al. (1995b):
the observed zero point was adopted for the F814W filter and the
synthetic zero point for the F450W and F606W filters.
To convert the F450W, F606W, F814W magnitudes into the B,V,I standard system 
we used the relation derived by Holtzman et al. (1995b) for the F606W filter 
and in Matthews et al. (1999) for the F450W filter.

\subsection{Host galaxy and polar ring morphology in the NIR and optical 
bands} 
\label{morphology}

\begin{figure}
\epsscale{0.5}
\plotone{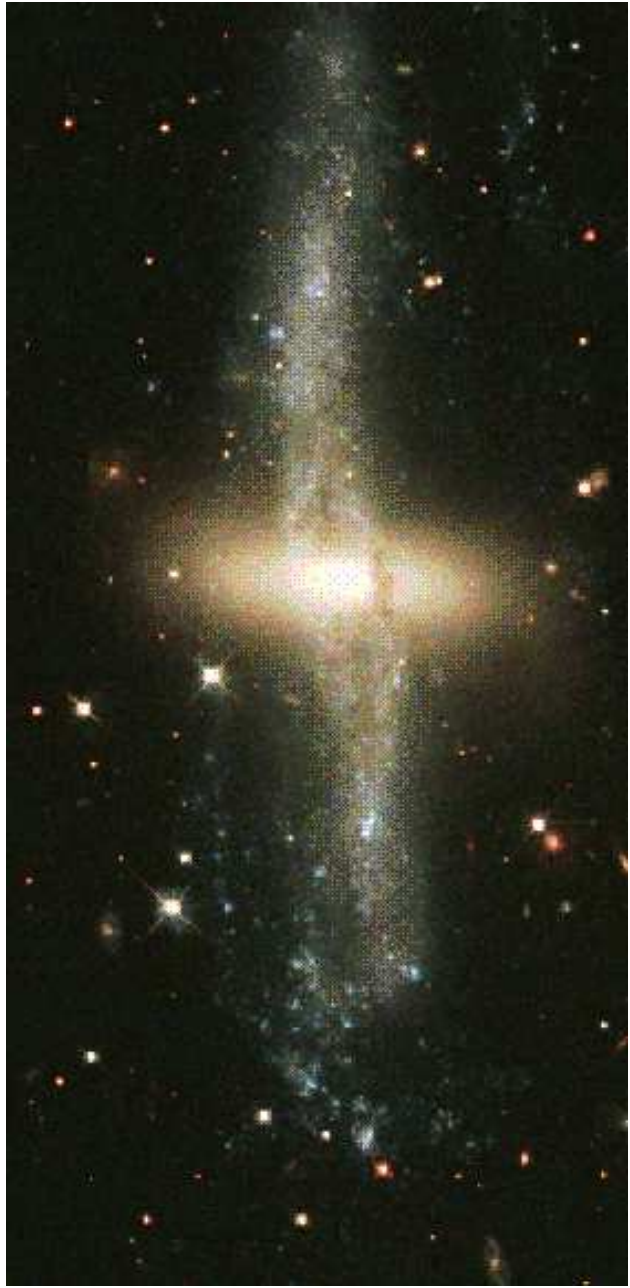}
\caption{\label{fig2} Color composite image of NGC~4650A from the
HST Heritage program. The image size is $1'.3 \times 2'.7$ . North is
$20^\circ$ counter clockwise from the y axis, East is $110^\circ$
counter clockwise from the same axis, on the left side of the image.}
\end{figure}

The J, H and Kn images of NGC~4650A show that the host galaxy is
the dominant luminous component, and its morphology resembles that 
of an early-type object, most likely an S0 (see Fig.~\ref{fig1}). 
The high angular resolution of the HST images allow a more detailed study of 
the inner regions of the polar ring and central spheroid. The light
distribution associated with the host galaxy shows a very concentrated
central component and a shallower, thicker envelope (see Fig.~\ref{fig2}). 
The bulge to disk ratio (B/D) for this system will be discussed in detail 
in the following sections.
The polar ring is more extended in the optical (B, V and I bands) than
in the NIR, and it appears knotty and dusty. The new HST data
reveal the presence of young blue star clusters describing arches out of
the main polar ring plane, along the SE and NW directions 
of the main light distribution (Fig.~\ref{fig2}, 
see also Gallagher et al. 2001).

\section{Photometry}\label{photo}
Integrated magnitudes and colors are computed for
the two main components of NGC~4650A, host galaxy and ring. 
The structural parameters are derived from the modeling of
the surface brightness distribution. 

\subsection{Total magnitudes}\label{colors}
\begin{figure}
\epsscale{0.8}
\plotone{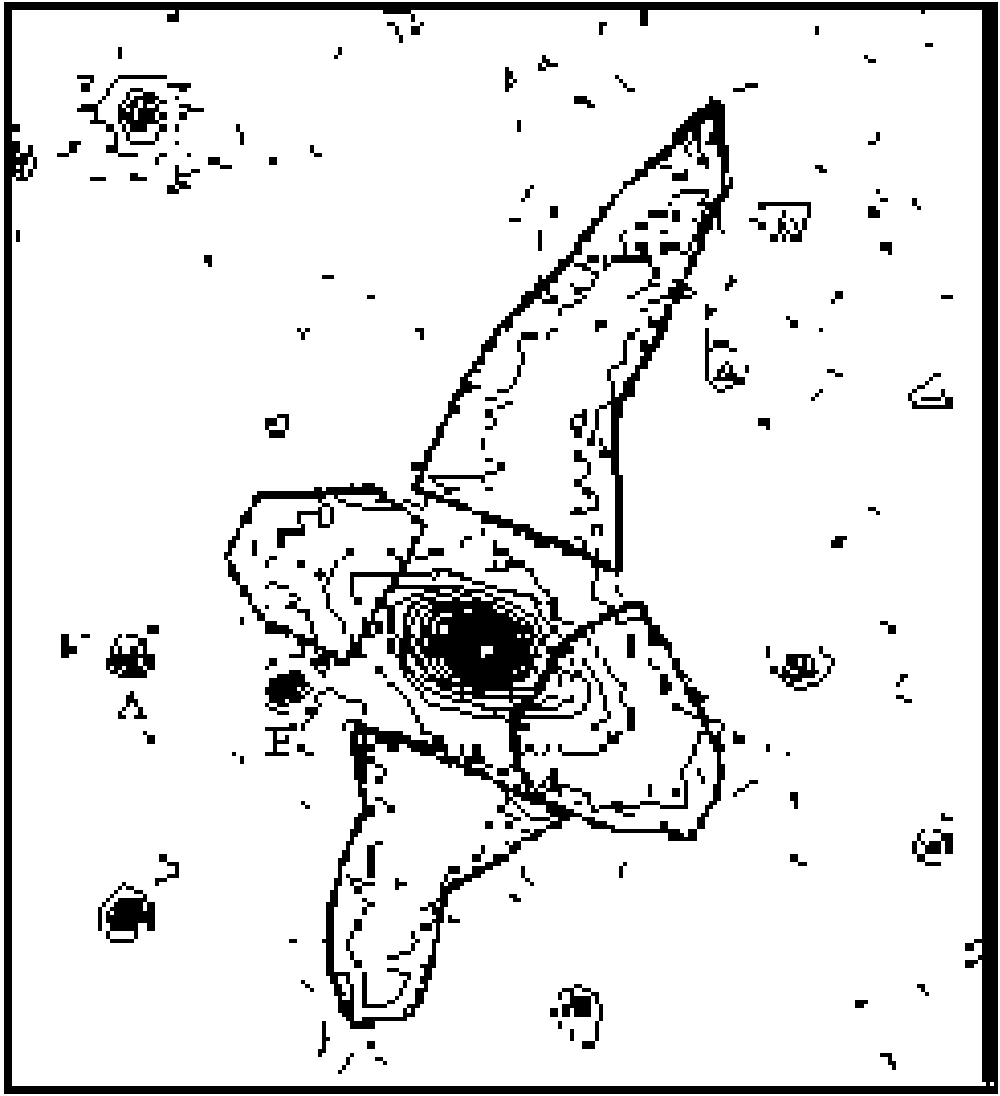}
\caption{\label{fig3} NGC~4650A contour plot in the J band plus
the 5 polygons limiting the different areas (heavier lines) where the integrated
magnitudes are computed. The distance between two labeled stars (A and B) is about
28 arcsec. North is up and East is to the left.}
\end{figure}

The integrated magnitudes are computed in B,V,I and J,H,Kn in five different areas. 
These areas are chosen as follows:
one is coincident with the nucleus; two areas, SW and NE of the nucleus, 
are placed within the host galaxy stellar component (outside the nucleus, 
in regions unperturbed by the polar ring),
and two areas for the polar ring, NW and SE of the galaxy center; 
see Fig.~\ref{fig3} for a complete summary.
The polygons are determined from the J image (using
the IRAF task POLYMARK) and used for all bands, 
after the images were registered and scaled to the J
image.
The integrated magnitudes inside each polygon are evaluated using the IRAF
task POLYPHOT. The photometric errors take into account both
statistics and background fluctuations. In the NIR bands, the error
relative to the integrated magnitude in each area is given 
multiplying the standard deviation per pixel (defined in Sec.\ref{NIRobs})
by the number of pixels in that area. The average error on integrated
magnitudes is about $0.06$ mag, which let to a 20\% error on colors. 
In the optical,
the error estimate includes also the uncertainties on the 
transformations to the standard B, V and I bands, and the average error 
on integrated magnitudes is about $0.04$ mag, which let to a 13\% 
error on colors.
The integrated magnitudes and colors correspondent to 
each area are listed in Table~\ref{tab3} and they have been corrected 
for extinction within the Milky Way, using
$A_{B}=0.485$ from Schlegel et al. (1998).

\subsection{Light distribution}\label{profiles}
\begin{figure}
\epsscale{1.0}
\plottwo{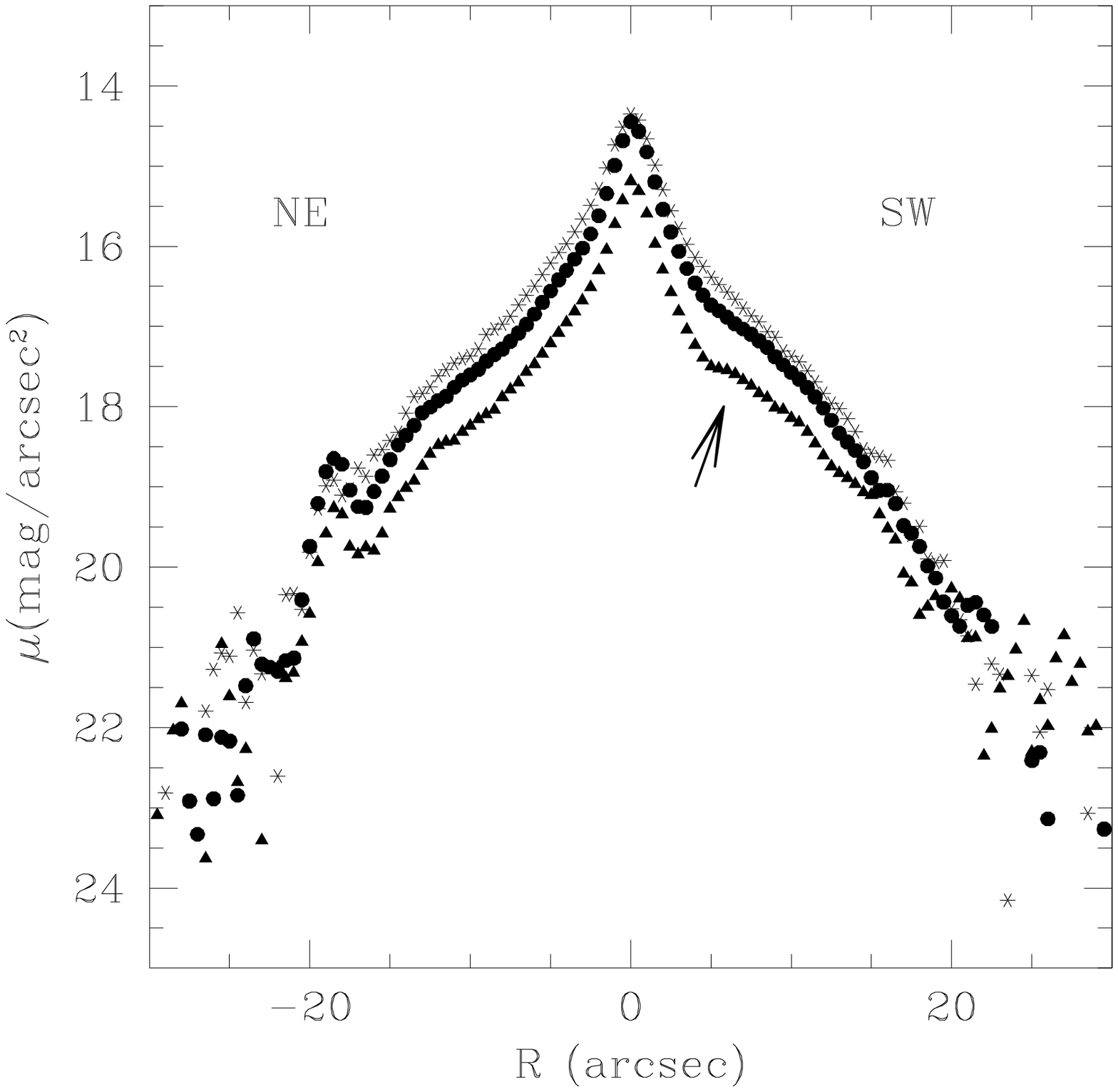}{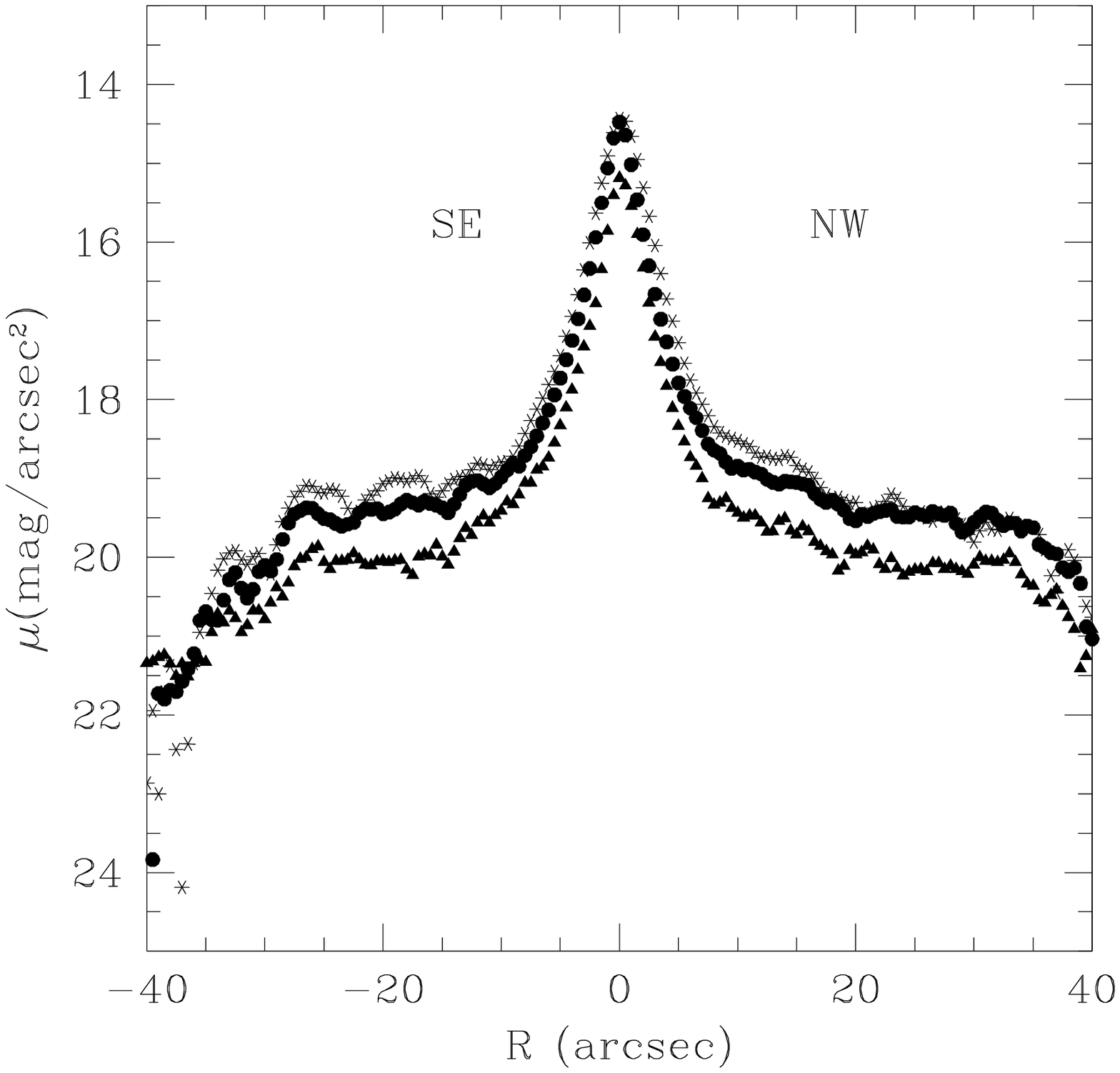}
\caption{\label{fig4} Left panel - NIR surface brightness profiles along the
S0 major axis,
at $P.A.=62^\circ$. The arrow indicates the absorption dip caused by
the polar ring on the central spheroid.
Right panel -  surface brightness profiles along the polar ring major axis, at
$P.A.=162^\circ$. The code for symbols is: J band full triangles, H band full
dots, and Kn band asterisks. }
\end{figure}

\begin{figure}
\epsscale{1.0}
\plottwo{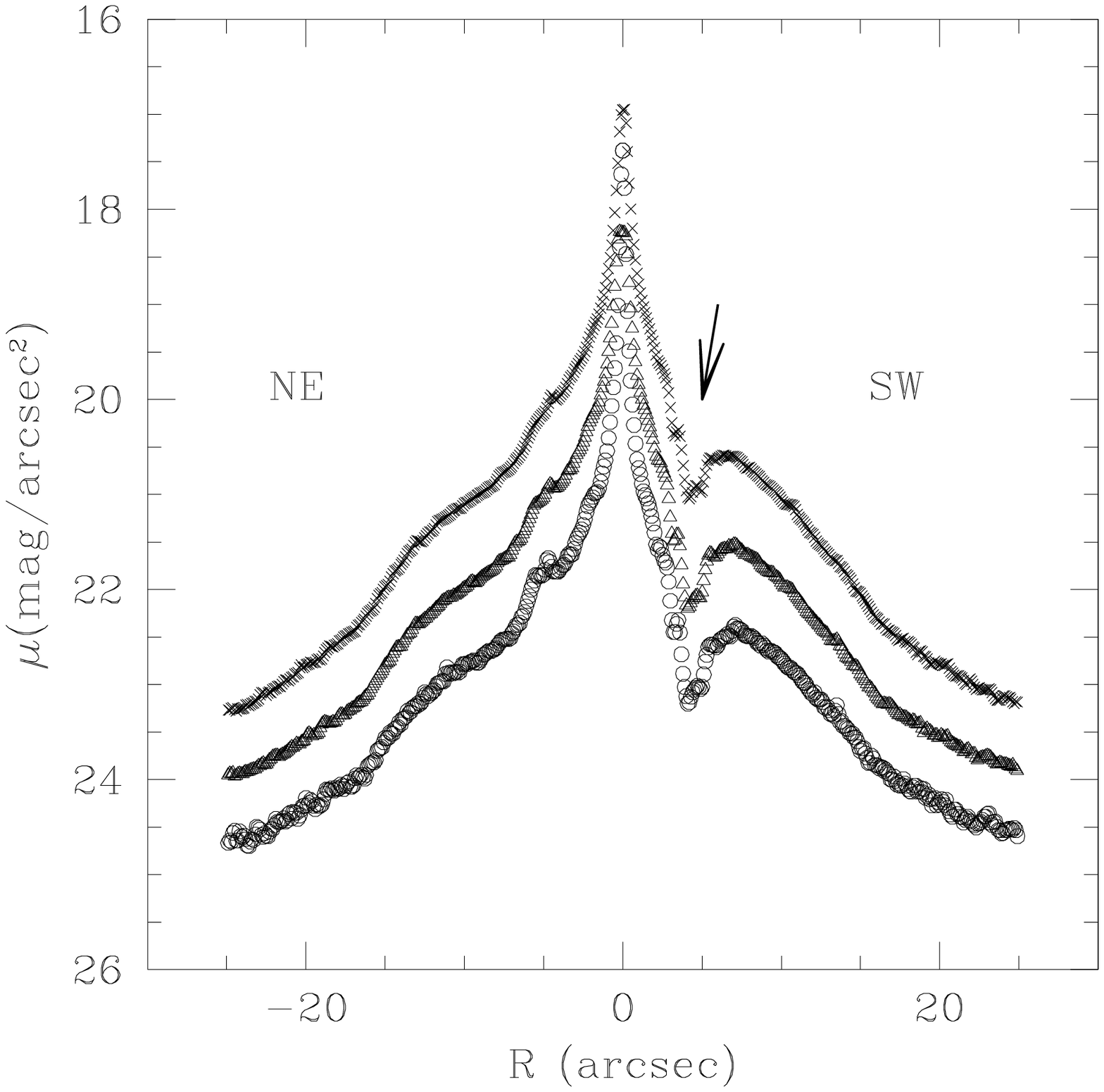}{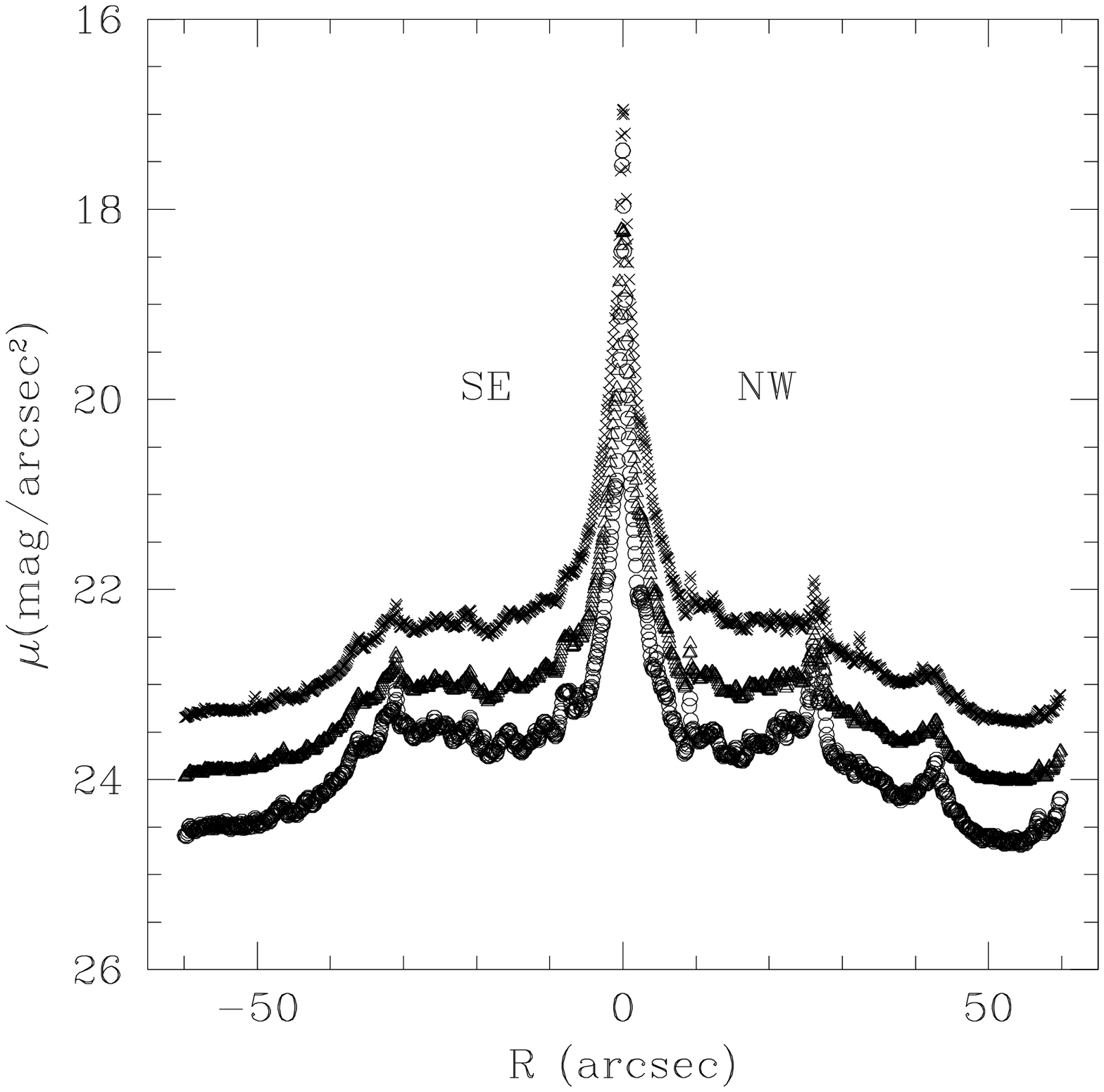}
\caption{\label{fig5}
Left panel - optical surface brightness
 profiles along the S0 major axis, at $P.A.=62^\circ$.
The arrow indicates the absorption dip caused by the polar ring on the central
spheroid.
Right panel - optical surface brightness profiles along the polar ring major axis, at
$P.A.=162^\circ$.
The code for symbols is: B band open dots, V band open triangles, and I
bands crosses.}
\end{figure}

In Fig.~\ref{fig4} and Fig.~\ref{fig5} the luminosity profiles are plotted 
for the major axis (left panels) of the S0 and polar structure 
(right panels), in the NIR (JHK) and optical (BVI) bands. 
The absorption due to the dust in the polar ring which passes in front of 
the central galaxy does perturb the J and optical profiles for this component 
at about 5 arcsec, SW of the galaxy center.
The polar ring shows a prominent {\it S-shape}, quite evident
in the NIR (Fig.~\ref{fig1}):  
because of this morphology, the polar ring light profiles 
are the result of an average of several profiles
extracted in a cone, centered on the S0 and 10 degree
wide ($\pm 5^{\circ}$ from the $P.A.=162^\circ$ of the ring major axis).
The presence of young star clusters, HII regions and dust is responsible 
for several bumps in the optical profiles (Fig.~\ref{fig5}), 
which appear less regular than those in the NIR (Fig.~\ref{fig4}).

\subsection{Two-dimensional model of the S0 light distribution}
\label{model1}
The 2D model of the S0 light distribution is done in the Kn and 
in the I band, because the  effect of dust absorption is weaker 
in these bands.
In the I band, the dust in the polar ring arm
does absorb the S0 light, as we outlined in Sec.\ref{profiles}.
The regions affected by this strong absorption are not taken into account
in the fitting routine: those symmetric with respect to the galaxy center were
used instead. Moreover the regions affected by 
foreground stars, and by the polar ring light along the S0 minor axis,  
are accurately masked before performing the fit to the
light distribution.
The light distribution of the S0 galaxy was modeled
through the super-position of a spheroidal central component and an
exponential disk (Iodice et al. 2001, Byun \& Freeman 1995). 
The projected light of the spheroidal component follows the generalized 
de Vaucouleurs law (Caon et al., 1993):
\begin{equation}
\mu_{b}(x,y)=\mu_{e}+k \left[\left(\frac{r_{b}}{r_{e}}\right)^{1/n}-1
\right]
\end{equation}
with $k=2.17n-0.355$, $r_{b}=\left[{x^{2}+y^{2}/q_b^{2}}\right]^{1/2}$, 
$q_{b}$, $\mu_{e}$ and $r_{e}$ are the
{\it apparent axial ratio}, the
{\it effective surface brightness} and the {\it effective radius}
respectively. 
The projected light distribution of the exponential disk (Freeman, 1970) 
is given by
\begin{equation}
\mu_{d}(x,y)=\mu_{0}+1.086\left(\frac{r_{d}}{r_{h}}\right)
\end{equation}
with $r_{d}=\left[{x^{2}+y^{2}/q_d^{2}}\right]^{1/2}$, $q_{d}$
$\mu_{0}$ and $r_{h}$ are the {\it apparent axial ratio},
{\it the central surface brightness} and the
{\it scalelength} of the disk respectively.

\begin{figure}
\epsscale{1.0}
\plottwo{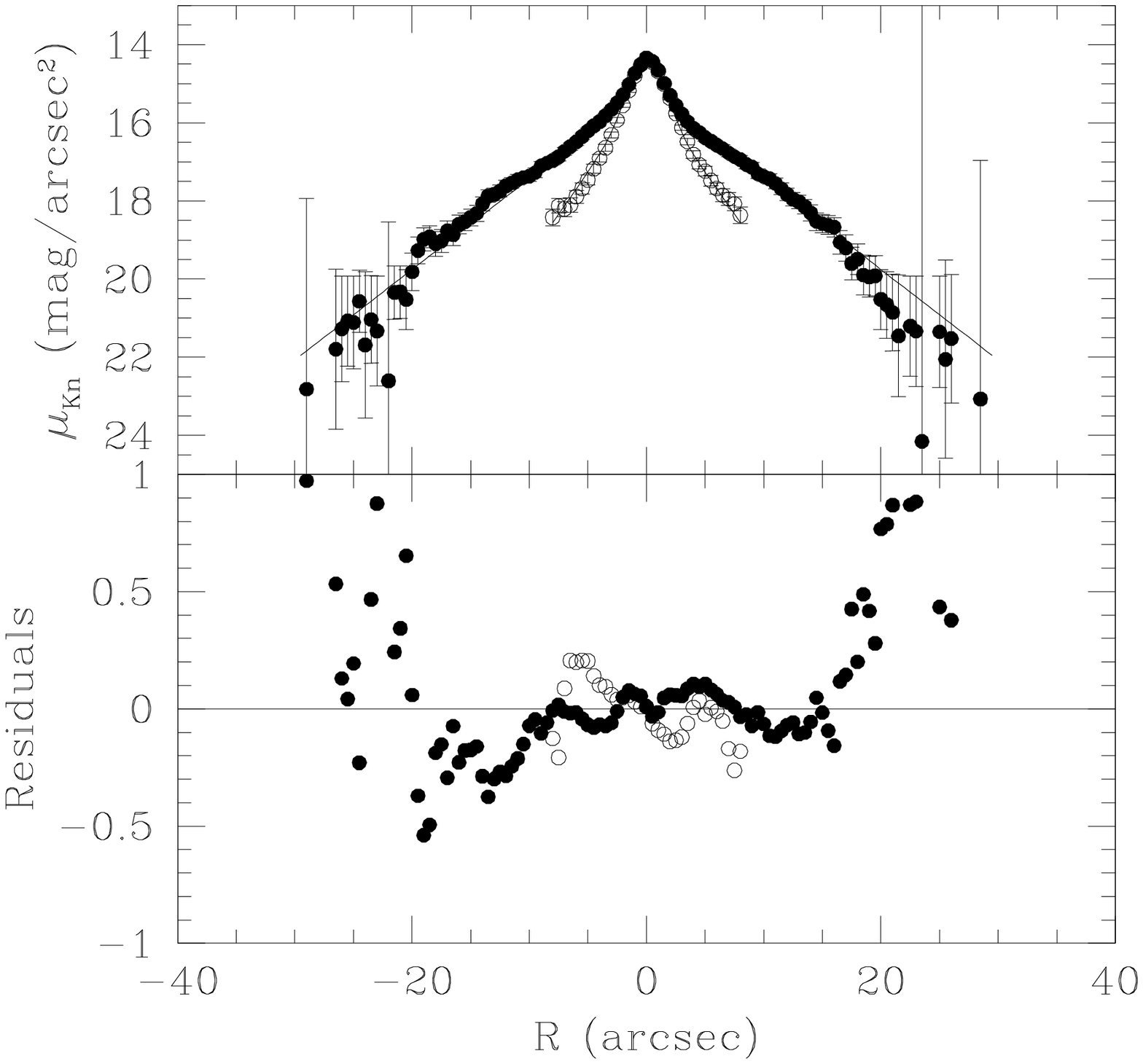}{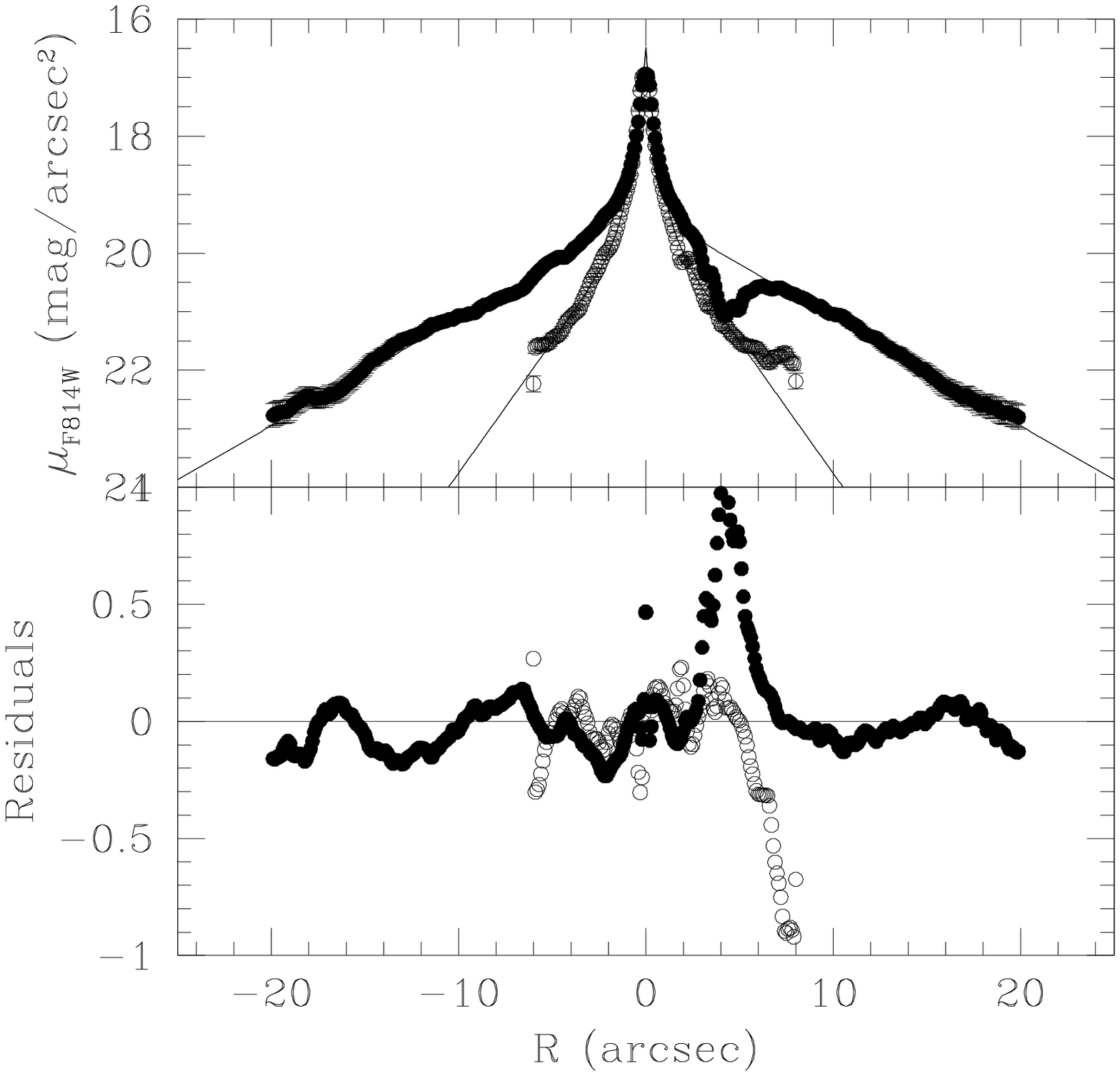}
\caption{\label{fig6}
Left panel - Fit of the S0 2D model to the light profile in
the Kn band to the major (filled dots) and minor axis (open dots).
Right panel - Fit of the S0 2D model to the light profile in the
I band for the major (filled dots) and minor axis (open dots).}
\end{figure}

\begin{figure}
\epsscale{0.8}
\plotone{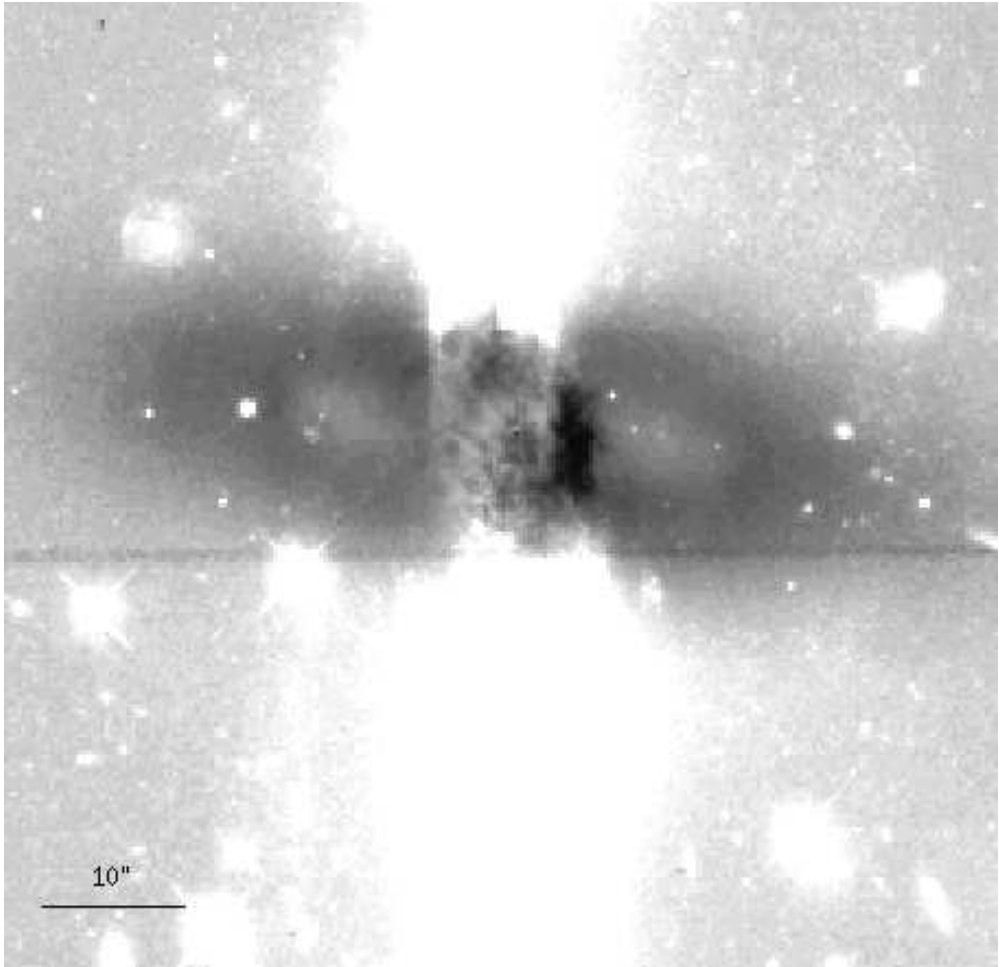}
\caption{\label{fig7}
Residual image obtained as the ratio between the whole galaxy frame and the
2D model for the S0 in V band.
Details on image processing  and 2D modeling of the light from the central
component are discussed in Sec.\ref{model1}.
Units are intensity; whiter colors correspond to those regions where the galaxy
is brighter than the model, darker colors corresponds to those regions where
the galaxy is fainter than the model.
North is $20^\circ$ counter clockwise from the y axis, East
is $110^\circ$ counter clockwise from the same axis,
on the left side of the image.}
\end{figure}

Fig.~\ref{fig6}  shows the comparison between the
observed and calculated light profiles in the Kn band (left panel) and in 
the I band (right panel). 
The structural parameters are listed in Table~\ref{tab4}. 
This model is fitted to the light
distribution of the S0 in such a way that there are no negative residuals,
i.e. this is the ``minimum'' model of the S0 light in the Kn and in the I
bands. The S0 model in the J and H bands, and in the B and V bands, are 
simply scaled versions of the Kn and I band models respectively,  
based on the average colors of the stellar component 
(see Sec.\ref{colors}).
The photometric errors do take into account the photon statistics
and background fluctuations (see Sec.\ref{NIRobs}). 
The effect of the Point Spread Function (PSF) was also considered in 
the central regions (by masking a region around the nucleus somewhat larger 
than the PSF). Fig.~\ref{fig7} shows
the ratio between the HST image for the whole galaxy and the S0 model, 
in the V band.
We see that the central galaxy is a stellar disk which has a warp in
the outer parts.  
The bright features near to the galaxy center connects
the outer parts of the polar ring and the nucleus of the system, this suggests
that there is no central hole, which is totally empty, at small radii 
in the polar ring light distribution. Similar results are 
obtained for the 2D modeling of the light distribution in the
NIR images. 
The values for the apparent $q_d$ ratio of the S0 disk, which is $\sim 0.5$,
should be considered an upper limit to the true flattening: 
the warp present in the S0 disk makes the isophotes more boxy in the
outer parts and therefore may produce a larger $q_d$ value.

\subsection{Scale parameters for the S0}
\label{params0}
The study of the host galaxy light distribution and its structural parameters
will help to understand whether this component is really a standard
early-type system. 
The optical light distribution in the host galaxy central regions has a 
quasi-exponential behavior: 
the value of the {\it n} exponent in Equation 2 is about 1, and 
the light is very concentrated toward the center, 
as the small value of the effective radius suggests. 
The convolution of this bright nuclear light concentration with the
PSF on ground based images may have caused a smoothing of the light
profiles toward the center, leading to a spuriously larger value for  $r_{e}$ 
and a smaller value for the {\it n} exponent in the NIR S0 model respect to the
I band one. 
The scale parameters for the central spheroid in NGC~4650A   
may be compared with those obtained by Caon et al. (1993) for a sample
of early-type galaxies in the Virgo cluster.  The $n$ exponents derived
by Caon et al. (1993) are those along the minor axis of the system, 
in order to exclude the contribution  from a possible disk component.
The typical range is $1 < n < 10.7$, while the
effective radius is in the range $0.44 < r_{e} \mbox{(kpc)} < 20.0$.
The bulge parameters for NGC~4650A
do not fall in the same area occupied by early-type galaxies in 
the $n$ - $r_{e}$ space: it is more centrally peaked, as indicated by
its small effective radius.
The B/D ratios in the I and Kn bands are very small: $B/D=0.107$ in the I band 
and $B/D=0.122$ in the Kn band; 
they are smaller than the typical values expected for
standard S0 galaxies for which $0.2 < B/D < 3.0$ (Bothun \& Gregg, 1990). 
The disk scalelength of the central spheroid in the I band 
is about 1 arcsec larger than in the Kn band,
which may suggest that there could be a difference in the
stellar population of the S0 disk as function of radius.

\subsection{Study of the light distribution in the polar structure}\label{parampr}
\begin{figure}
\epsscale{1.0}
\plottwo{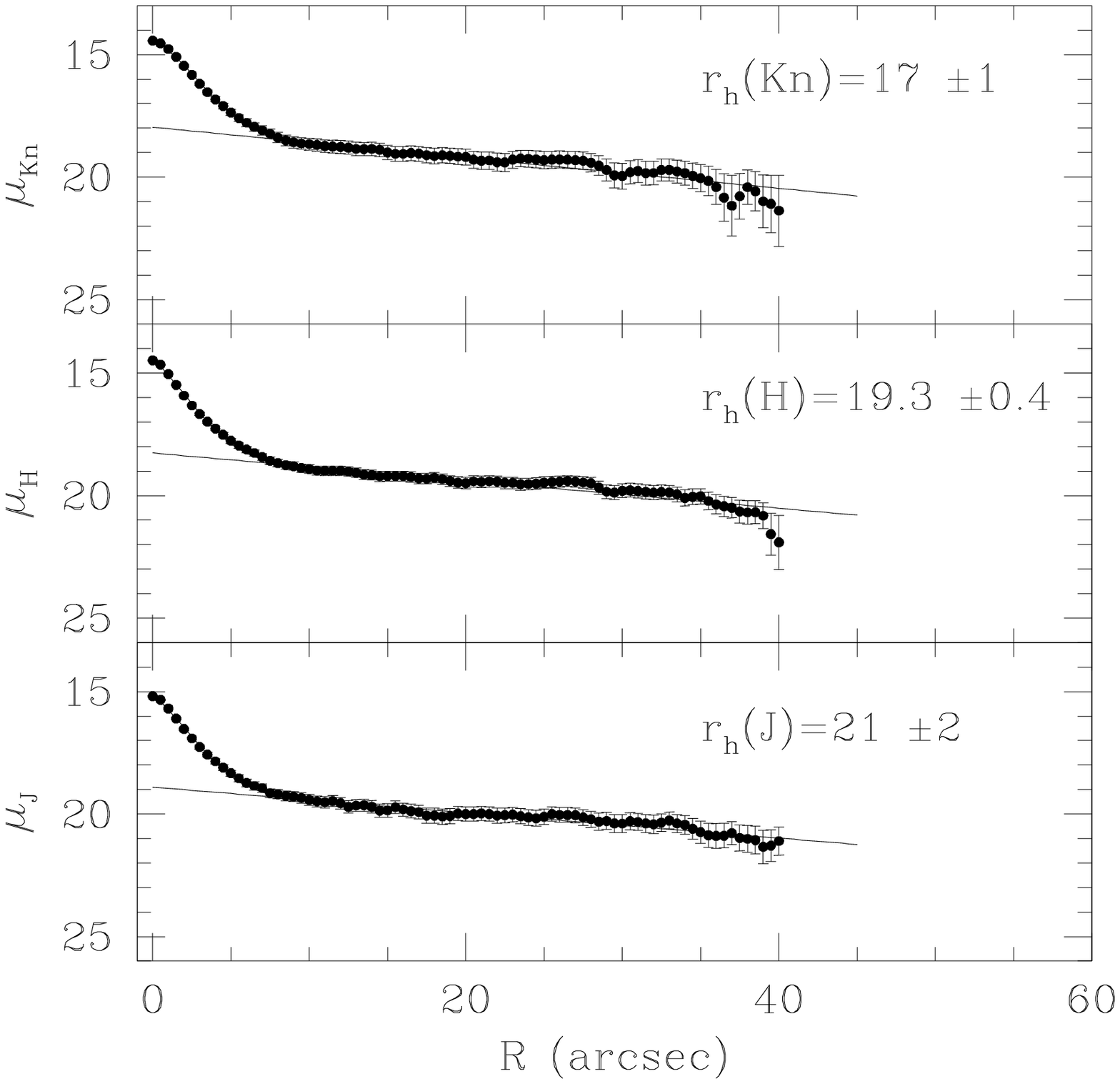}{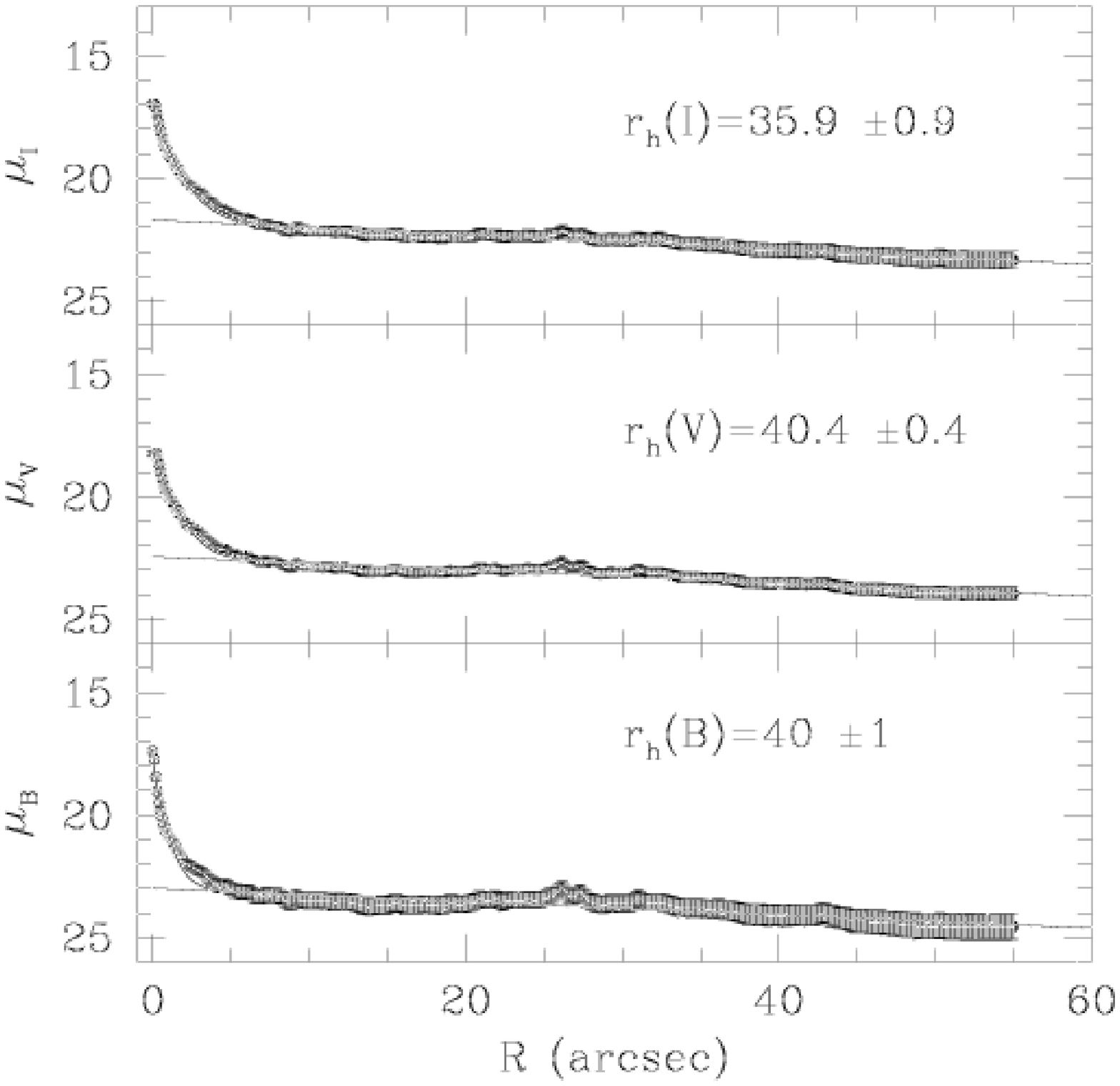}
\caption{\label{fig8} Left panels: average luminosity profiles for the
polar ring in J (bottom), H (middle) and Kn band (top).
Right panels: average luminosity profiles for the polar
ring in B (bottom), V (middle) and I band (top).
The best fitting exponential is overlaied on the data points in each plot;
scalelengths are reported in arcsec. }
\end{figure}

Because of the polar ring morphology in the optical, the luminosity profile
along the polar ring major axis is computed as averages of 20
extracted profiles, parallel to P.A. $= 162^\circ$, and
the final profile is obtained as average of the two sides
opposite to the nucleus.
The polar ring light profiles are well reproduced by an exponential law and
the comparison between the observed average profiles and the relative best
fit is shown in Fig.~\ref{fig8}. The polar ring scalelength 
decreases from the optical to the NIR bands, i.e. the polar ring is
more extended in the optical than in the NIR, and
may also suggests the presence of different stellar
populations: an older inner component and
a more extended, younger, one.

An important quantity related to the size of the polar ring 
is the moment of its radial distribution 
\begin{equation}
(\Delta R)^2 = \frac{\int_{r_{e}}^{\infty}(r-\bar{R})^{2}*\mu(r) dr}{\int_{r_{e}}^{\infty} \mu(r) dr}
\end{equation}

where $\bar{R}$ is 
the average radius 
\begin{equation}
\bar{R} = \frac{\int_{r_{e}}^{\infty} r*\mu(r) dr}{\int_{r_{e}}^{\infty} \mu(r) dr}
\end{equation}
weighted by the surface brightness distribution and $r_{e}$
is the effective radius of the central component. 
For a pure exponential disk, the  $\Delta R / \bar{R}$ ratio tends to 
unity when {\it r} tends to infinity.
For a real object, this value is expected to be less than 1, because of its 
finite extension.
This is confirmed by the  $\Delta R / \bar{R}$ values derived for a sample of 
spiral galaxies (de Jong 1996) in the B band: this quantity varies 
from $45\%$ to $75\%$ and the average value is $\Delta R /\bar{R} \sim 65\%$. 
In the B band, the polar structure in NGC~4650A and spiral galaxies 
in the de Jong sample have similar scalelengths  
(for spirals $13 \le r_h \mbox{(arcsec)} \le 65$, 
with an average value of about $25 \pm 11$ arcsec). 
The polar ring is less luminous than typical spirals (see
Table~\ref{tab3}),
so the $\Delta R / \bar{R}$ for the polar ring
is going to be smaller than the average value obtained for spiral galaxies.
This value is computed for the I band (which is less disturbed by dust 
absorption) and gives $\Delta R / \bar{R}\sim 50\%$.
The structural parameters of the light distribution and its extension
suggest that the polar structure is more similar to a disk than
an annular ring.

\section{Using colors to date the stellar populations of 
NGC~4650A}\label{stellar}
\begin{figure}
\epsscale{1.0}
\plottwo{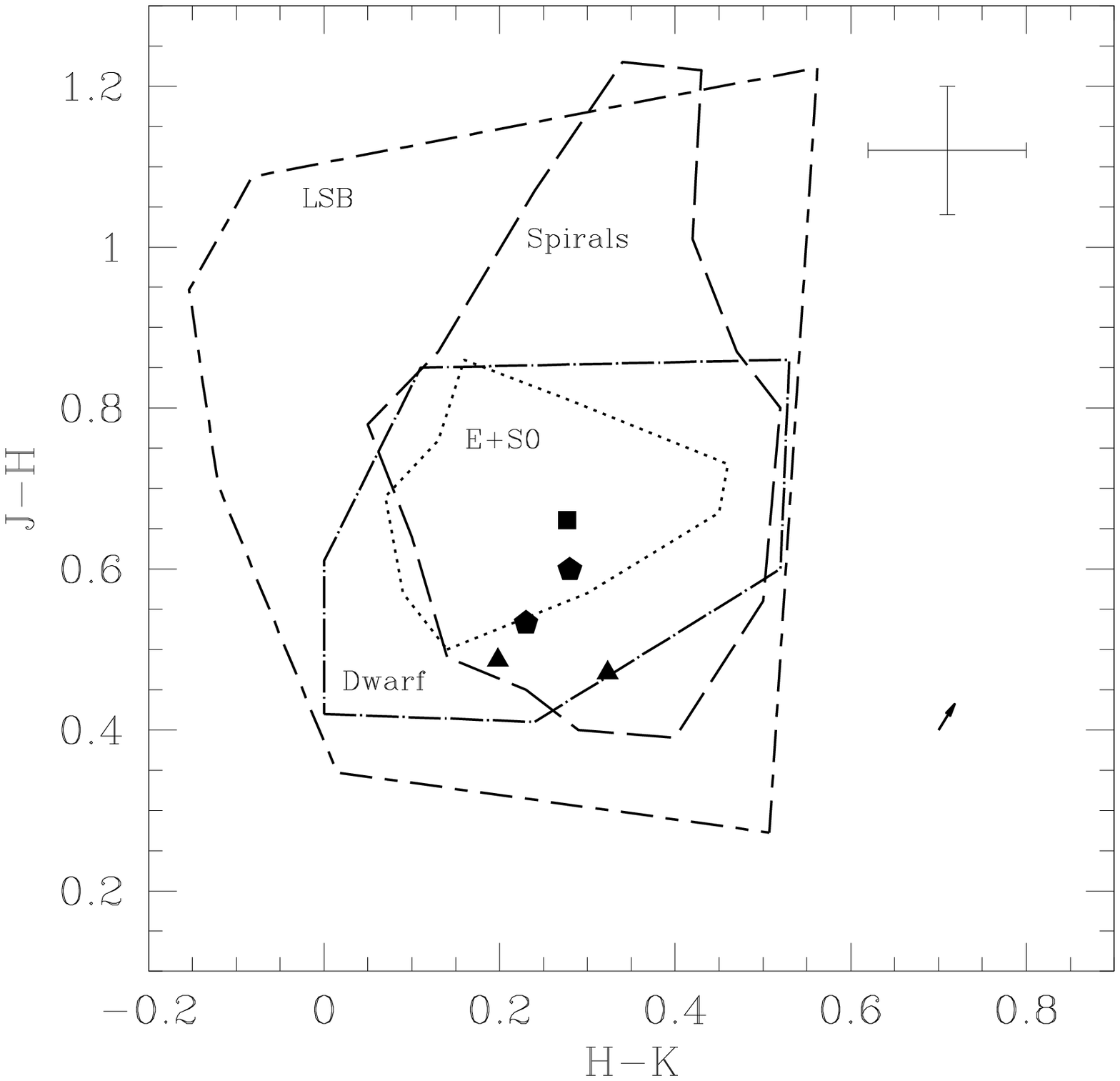}{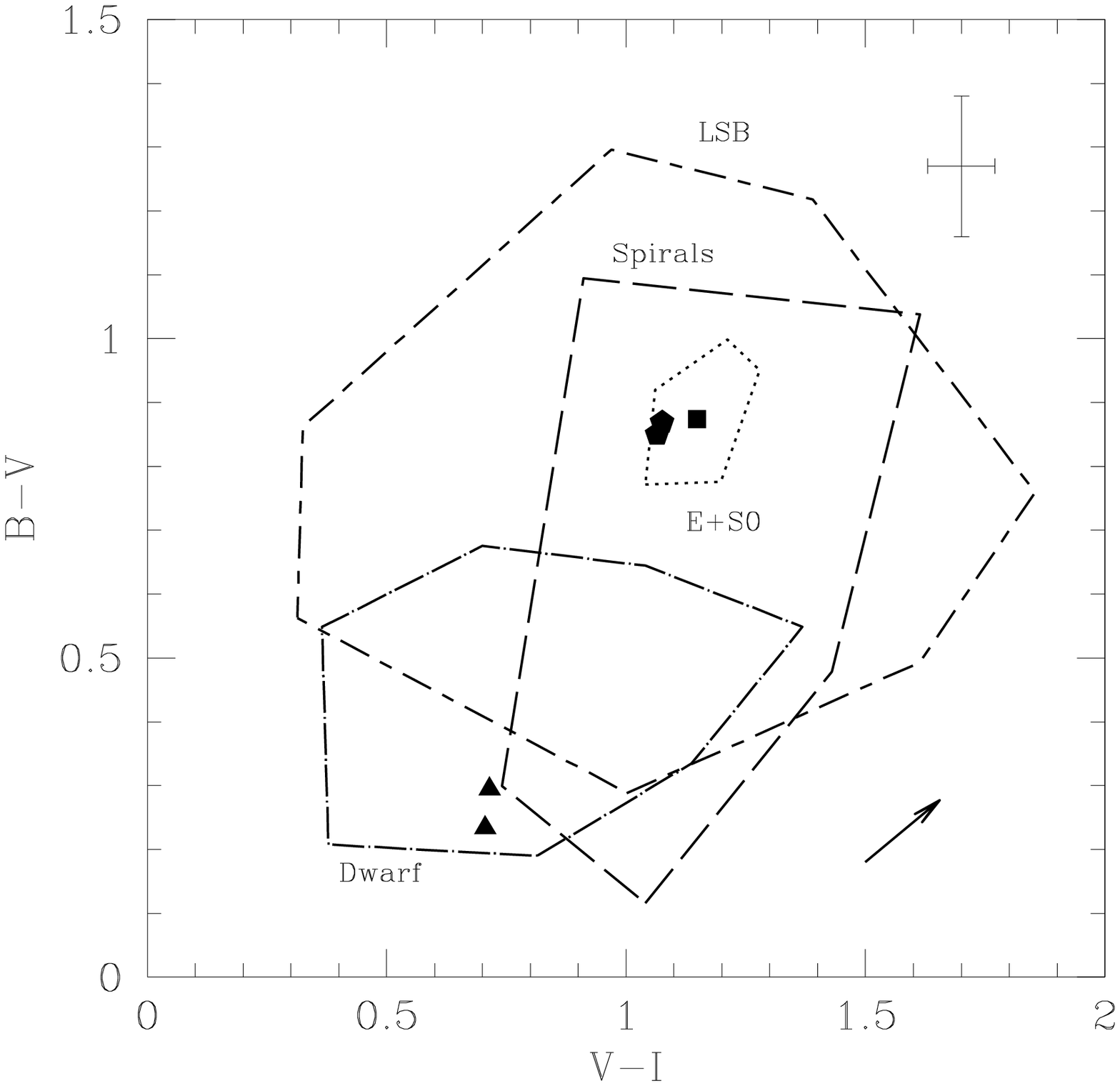}
\caption{\label{fig9}
J-H vs. H-K (left panel) and B-V vs. V-I (right panel) color diagrams for the
five areas of the NGC~4650A: filled square indicates
the central region of the host galaxy; filled pentagons indicate the
stellar component outside the central region; filled triangles indicate the polar
ring regions.
The dotted contour limits the region where the Es and S0s integrated colors are
found; the long-dashed contour limits the integrated colors of spirals; the
dashed-dotted contour identifies the integrated colors of the dwarf galaxies
and the long dashed - short dashed
contour indentifies the integrated colors of LSB galaxies.
The arrow, in the lower right corner, indicates the reddening vector 
for galactic
dust and the screen model approximation, quoted in Sec.\ref{stellar}. 
The average errors on colors are showed in the top right corner.}
\end{figure}

One wishes to compare the integrated colors of the main components 
(host galaxy and polar structure, shown in Table~\ref{tab3}) 
with those of standard morphological galaxy types, and check 
whether differences in colors are related to dust absorption or 
to different stellar populations.
\noindent
{\it NIR colors -} The NIR colors of the central spheroid and the polar 
structure are compared 
with those of (1) standard early-type galaxies in the Fornax and Virgo 
clusters (Persson et al., 1979), (2) spirals (Giovanardi \& Hunt, 1996; 
Frogel, 1985; de Jong \& van der Kruit,
1994), (3) dwarf ellipticals (Thuan, 1985), and (4) low surface brightness 
galaxies (Bergvall et al. 1999).
The J-H vs. H-K plot in Fig.~\ref{fig9} (left panel) shows that the nucleus 
of the central 
component is redder than its outer regions and falls in the area 
occupied by early-type galaxies. The other two regions
show bluer colors and are close to the area identified by the dwarf galaxies.
The color gradient through the host galaxy spheroid may be accounted for by
the reddening due to the dust, as indicated by the
reddening vector computed for a screen model approximation
and $A_{V}=0.3$ (Gordon, Calzetti \& Witt 1997).
On average, the polar ring is bluer than the central galaxy. 
There is an additional 
color difference between the two regions of the polar ring: 
the South side of the polar structure has 
a redder H-K color with respect to the North side, 
which cannot be accounted for by the reddening vector alone.
It may be caused by a different dust distribution and high scattering
in the two regions of the polar ring: the screen model is inadequate to describe it. 

\noindent
{\it Optical colors -}The B-V vs. V-I colors for the S0 component and the 
polar ring (Fig.~\ref{fig9}, right panel) are compared with those of (1) standard 
early-type galaxies (Michard \& Poulain 2000), (2) spiral galaxies 
(de Jong \& van der Kruit, 1994),
(3) dwarf galaxies (Makarova  1999),
(4) LSB galaxies (O' Neil et al. 1997; Bell et al. 2000). 
The optical colors of the host galaxy appear very similar to those of 
early-type objects.
If one accounts for a reddening caused by the dust in the polar ring,
in the screen model approximation, 
the colors of the central spheroid will fall in the region for
late-type systems. The reddening due to dust can also account for the 
color gradient between the center and outer regions along the polar 
ring major axis. On the other hand, the difference between the integrated 
colors of the polar structure and those of
the central component cannot be accounted for by the 
reddening vector alone: such a large gradient is more likely due to a difference 
in stellar populations.

\begin{figure}
\epsscale{1.0}
\plottwo{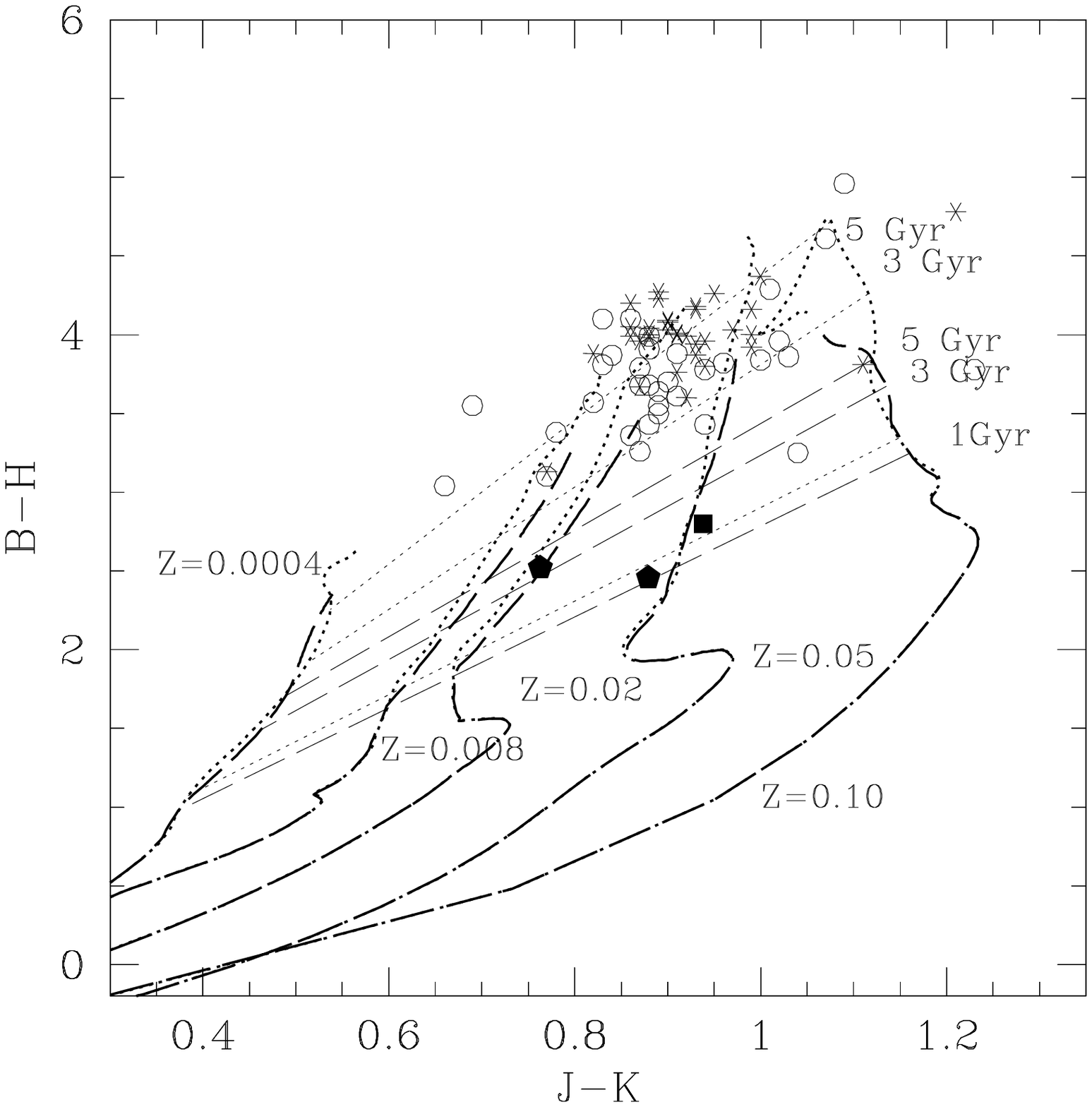}{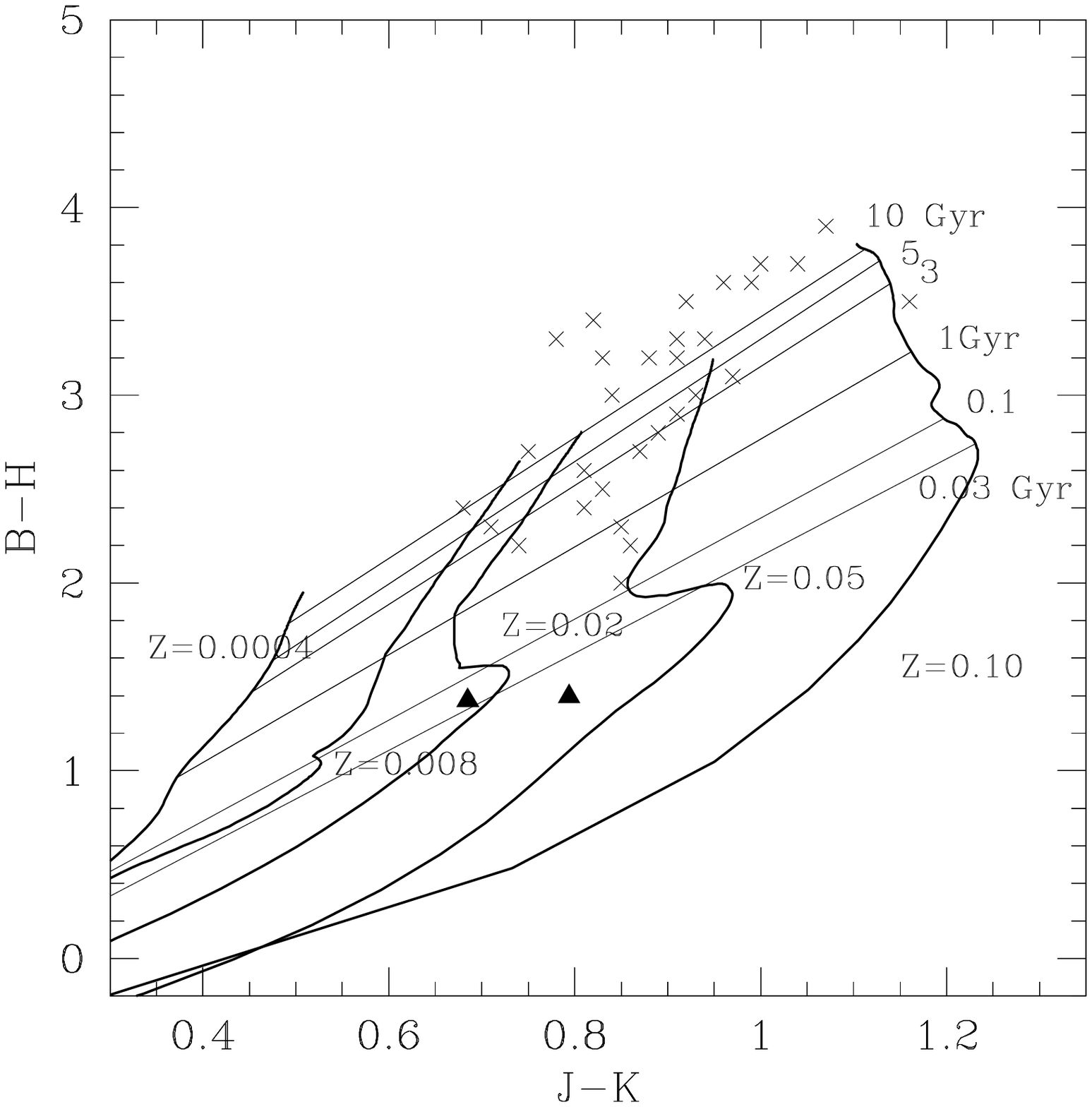}
\caption{\label{fig10} B-H vs. J-K diagram of the evolutionary tracks
for the stellar synthesis models optimised for
the central component (left panel) and the polar-ring component (right
panel).
Left panel: the heavier dotted lines correspond to models with a characteristic
timescale $\tau=1 Gyr$ and heavier dashed lines for models with $\tau=7 Gyr$.
Models are computed for
different metallicities as shown on this figure.
Light dotted and light dashed lines indicate loci of constant age for the
different models;
different ages are reported on the plot. 
The filled square and pentagons correspond
respectively to the nucleus and the outer regions of the central spheroid 
in NGC~4650A,
open circles and asterisks correspond to bulges and disks from a sample
of S$0$ galaxies (Bothun, 1990).
Right panel: heavier lines indicate model with constant SFR computed for
different metallicities (as reported on the plot). Light lines are loci of
constant age; different ages are quoted on the plot.
Filled triangles indicate the polar ring regions;
crosses are for a sample of spiral galaxies (Bothun et al., 1984).}
\end{figure}

The stellar population synthesis model by Bruzual \& Charlot (1993) 
were used to reproduce the integrated colors of different regions 
(see Sec.\ref{colors}) in the polar ring galaxy NGC~4650A. The goal is to 
derive an estimate of the stellar population ages in the central spheroid 
and the polar structure.
As a first step, we selected a set of models which were able to reproduce 
the average integrated colors of galaxies with different morphological 
types in the local Universe, and then they were optimized to reproduce the 
colors observed for the two main components of NGC~4650A, in particular
the B-H and J-K colors. 
The B-H vs. J-K diagram is used to break the age-metallicity degeneracy, 
as suggested by Bothun et al. (1984). The J-K color is a good estimate 
of the metallicity and it is quite insensitive to the presence of a young 
stellar population. This is supported by the observed monotonic increase 
of the mean J-K color in globular clusters with increasing metallicity 
(Aaronson et al. 1978, Frogel et al. 1983), and the 
population synthesis models by Bothun (1982) show that
J-K is decreased only by $0.05$ mag as a result of a starburst,
while the B luminosity is increased by $1$ mag. On the other hand,
the B-H color is sensitive to the combined effect of SFR, metallicity and 
age (Bothun et al. 1984).
Fig.~\ref{fig10} shows that the central component is
overall bluer in the B-H color than the average values for early-type galaxies, so
a younger age is to be expected. 
The polar structure is significantly bluer than the central galaxy, 
implying even a younger age for its stellar population. 

A star formation history with an exponentially decreasing rate was
adopted for the central component. It has 
the following analytical expression: 
$SFR(t)= \frac{1}{\tau} \exp{(- t/ \tau)}$, where
the $\tau$ parameter quantifies the ``time scale'' when the
star formation was most efficient. Adopting $\tau=1$ Gyr and $\tau =7$ Gyr,
the correspondent evolutionary tracks were derived for different 
metallicities ($Z=0.0004$, $Z=0.008$, $Z=0.02$, $Z=0.05$, $Z=0.1$), 
which were assumed constant with age.  
As shown in the left panel of Fig.~\ref{fig10}, 
these models reproduce the photometric properties of early-type 
galaxies in the local Universe.
A constant star formation model (with metallicities: 
$Z=0.0004$, $Z=0.008$, $Z=0.02$, $Z=0.05$, 
$Z=0.1$) which reproduces the integrated colors of local spiral galaxies 
(right panel of Fig.~ \ref{fig10}) was used for the polar structure.
In every model it has been assumed that stars form according to the
Salpeter (1955) IMF, in the range from $0.1$ to $125 M_\odot$.                  
The lines of constant age were computed from the evolutionary tracks and
suggest an age between $1$ to $3$ Gyr for the central component 
(left panel of Fig.~ \ref{fig10}),
has an overall younger age than the typical ages of an early-type system. 
The age derived for the polar structure is less than $10^8$ yr  
(see right panel of Fig.~ \ref{fig10}).
A cautionary note: our derived colors are all upper limits, since they 
were not corrected for the absorption caused by the dust in the polar
structure and indeed the true colors of the central stellar component
might be even bluer. Furthermore, the age estimates for the central component and
the polar structure are uncertain because we lack
independent information on the star formation law and metallicity
of the stellar population in the central host galaxy and polar structure.
The intrinsic uncertainties of the synthesis population models must
also be considered, particularly for the age of the central component.
By comparing three recent synthesis codes, Charlot, Worthey and
Bressan (1996) found that the colors predicted for old populations with
an age $>1$ Gyr,
plus same input age and metallicity, are affected by
discrepancies, which are primarily due to the different prescriptions  
adopted for the stellar evolution theory.
Thus, our estimates are only indicative of the relative ages between the
central host galaxy and polar structure.

\section{Constraints on the formation scenario for NGC~4650A: 
conclusions }\label{formation}

We have presented a detailed photometric study of 
polar ring galaxy NGC~4650A, based on new NIR observations and
high resolution optical imaging acquired with the HST. 
We now wish to compare our results against
the properties predicted for PRGs in different formation scenarios.
Possible scenarios for polar ring formation can be grouped into two main 
pictures, i.e  1) accretion of
a gas-rich dwarf galaxy by an early-type system or gas stripped from a nearby 
gas-rich object, and 2) a major dissipative merging of two disk galaxies.
Accretion in, or gas-stripping by, an oblate/triaxial galaxy
can predict the formation of a narrow polar annulus. 
These annuli can be in a quasi-equilibrium 
configuration if 1) their ratio $\Delta R / \bar R$ is between $10\%$ and $30\%$, 
where $\Delta R$ is the radial extension of the ring and $\bar R$ the 
average radius (Katz \& Rix, 1992; Christodoulou et al. 1992; 
Reshetnikov \& Sotnikova, 1997) 
and 2) self-gravity is at work (Sparke 1986, Arnaboldi \& Sparke 1994).
The total amount of accreted gas can be of the order of $10^{9} M_\odot$, 
which is the typical amount of HI in a gas-rich dwarf (Carignan 1999).
The process of ring formation through accretion/stripping 
of a gas-rich companion takes few Gyrs at most. In this scenario
the host galaxy is an early-type gas-free system and preserves its 
structure (luminosity profile, B/D ratio, scale parameters), while some 
star formation is triggered by the event (Arnaboldi et al. 1993).

The merging scenario of two disks can account for
the formation of narrow rings as well as of wide/massive disk-like
structures, depending on the initial mass ratio of the two merging
progenitors. According to this scenario, 
the polar structure represents what remains of the ``victim''
disk galaxy, while the accreted ``intruder'' has supplied the S0-like 
component.
Polar rings are more likely to form when the two disks 
encounter each other on a polar orbit,
with a small initial orbital angular momentum (Bekki 1998). 
This dissipative merger event transforms the intruder's thin stellar disk 
into a thicker structure. Both intruder's and 
victim's disk radial mass distribution deviates from the initial exponential 
profiles as the merger process goes on.
When extended polar structures are formed (depending on the 
intruder/victim mass ratio), they are characterized 
by the absence of a hole in the central regions, and their 
HI content can be as large as that of late-type 
spirals, i.e. up to $10^{10} M_\odot$.  
The predicted evolution time for the whole resulting polar ring galaxy is of 
about $10^{9}$ yr,  and the outwardly propagating gaseous waves, 
excited by the intruder galaxy, trigger a burst of stellar formation 
in the polar disk.

What are the observational properties of the polar ring galaxy NGC~4650A?
Can they discriminate between the two scenarios? 
Both in the NIR and HST images, we have found that stellar light and dust 
lanes connect the outer parts of the polar structure with the inner central 
nucleus of the system, i.e. this component does not
show a completely empty central hole. The light distribution of the polar 
structure follows an exponential profile and has a 
$\Delta R / \bar R \sim 50\%$. 
These properties suggest that this structure is more similar to a disk rather 
than a narrow annulus, as already suggested by Arnaboldi et al. (1997) 
through the analysis of the HI distribution in the polar structure. 
Furthermore,
the HI observations for NGC~4650A showed that the total HI mass in this
component is about $10^{10} M_\odot$.
The high resolution HST images indicate that the central spheroidal
component is not a standard early-type system. 
It has a small exponential bulge and a 
disk with an exponential profile, which appears slightly warped in 
the outer regions. This warp is observed both in the optical and in NIR.
The study of the integrated colors of NGC~4650A has shown that polar structure 
is much bluer than the stellar central component, as
it is also visible in the beautiful
HST multicolor images. The comparison with the predicted 
colors from stellar population synthesis codes suggests a very young 
age ($<0.5$ Gyr) for the polar structure. The age of the central 
component is in the range of $1$ to $3$ Gyr, which is of a
significantly younger age than those typical for early-type galaxies.

Whatever event may have occurred in the past of NGC~4650A, it has strongly 
changed the properties of the host galaxy, both in the structure and stellar 
population, so that this component differs from a standard S0 system.
Published simulations of the accretion/stripping scenario 
were not able to reproduce either these observed properties for the host galaxy in 
NGC~4650A, or those for the polar structure.
In particular, we note the absence of a hole in the center of the polar disk,
which has an exponential light distribution and a large amount of HI, which is
an order of magnitude larger than what is expected in the accretion models.
Furthermore, for reasonable mass-to-light ratios ($M/L \sim 2$ in the NIR, 
from Matthews, van Driel and Gallagher 1998), 
the luminous mass (gas plus stars) in the polar structure, of about 
$12 \times 10^{9} M_\odot$, is comparable with or even higher than the total 
luminous mass in the host galaxy, of about $5 \times 10^{9} M_\odot$.
In the accretion scenario, one would expect the accreted baryonic mass (stellar + gas)
to be a fraction of that in the pre-existing galaxy, and not viceversa,
as it is observed for NGC 4650A. 
Based on these new observational results, we have considered the alternative ideas
discussed by Bekki according to which the polar ring galaxies, 
such as NGC~4650A, may have formed from a dissipational merger event of 
some kind. The dissipative merger scenario proposed by Bekki (1998) may 
provide a coherent explanation for the wide and massive polar disk of NGC~4650A
and its larger baryonic mass content with respect to the central component,
plus the non-standard properties of the light distribution in the 
central S0-like component.
A future test of Bekki's scenario for PRG formation should also include
a detailed analysis of the kinematics predicted from the N-body simulations
and a comparison with PRG kinematics, and address the question about the stability 
of those merger configurations which lead to the formation of a massive disk.


\acknowledgments
The authors wish like to thank the anonymous referee whose comments and
suggestions greatly improved the presentation of this work.
E.I. and M.A. gratefully acknowledge the University of 
Wisconsin-Madison and the staff of the Astronomy department for 
their support. LSS acknowledges grant AST-9803114 from the
U.S. National Science Foundation.  JSG is grateful to the Graduate
School of UW-Madison and the Vilas Foundation for financial support
through a Vilas Associateship.

\newpage

\newpage

\begin{table}
\begin{center}
\caption{\label{tab1} NIR observation log of NGC~4650A .}
\begin{tabular}{ccccc}
\tableline\tableline
Filter & Tot. int. (s) & FWHM ('') & Image field & Date \\
\tableline
J & 1200 & 1.3 & 1 & 21/03/1995\\
J & 1200 & 1.3 & 2 & 21/03/1995\\
H & 1200 & 1.5 & 1 & 20/03/1995\\
H & 1200 & 1.4 & 2 & 20-21/03/1995\\
Kn & 2400 & 1.4 & 1 & 20/03/1995 \\
Kn & 2400 & 1.4 & 2 & 20/03/1995 \\
\tableline
\end{tabular}
\end{center}
\end{table}

\newpage

\begin{table}
\begin{center}
\caption{\label{tab2} HST observation log of NGC~4650A .}
\begin{tabular}{ccccc}
\tableline\tableline
Filter & Tot. int. (s) & FWHM ('') & n. exp. & Date \\
\tableline
F450W & 7500 & 0.24 &8 & 04/1999\\
F606W & 4900 & 0.31 &6 & 04/1999\\
F814W & 7600 & 0.28 &8 & 04/1999 \\
\tableline
\end{tabular}
\end{center}
\end{table}

\clearpage
\newpage

\begin{table}
\begin{center}
\caption{\label{tab3} Integrated magnitudes and colors of different regions
of S0 and of polar ring.}
\begin{tabular}{cccccccccccc}
\tableline\tableline
Component & Region & $m_{B}$ & $M_{B}$ & $m_{J}$ & $M_{J}$ & B-V & V-I & B-H & J-K & J-H & H-K \\
\tableline
PR & SE & 16.38 & -16.7 & 15.46 & -17.6 & 0.29 & 0.71 & 1.39 & 0.79 & 0.47 & 0.32 \\
PR & NW & 15.90 & -17.2 & 15.01 & -18.0 & 0.23 & 0.71 & 1.37 & 0.68 & 0.49 & 0.20 \\
S0 & SW & 16.72 & -16.3 & 14.73 & -18.3 & 0.87 & 1.07 & 2.52 & 0.76 & 0.53 & 0.23 \\
S0 & NE & 17.21 & -15.8 & 15.35 & -17.7 & 0.85 & 1.06 & 2.46 & 0.88 & 0.60 & 0.28 \\
S0 & center & 15.40 & -17.6 & 13.30 & -19.8 & 0.87 & 1.15 & 2.80 & 0.94 & 0.66 & 0.28 \\
\tableline
\end{tabular}
\end{center}
\end{table}   

\newpage
\clearpage

\begin{table}
\begin{center}
\caption{\label{tab4} Structural parameters for the host galaxy
in NGC~4650A.  The effective surface brightness $\mu_{e}$ and the central
surface brightness $\mu_{0}$ are in $mag/arcsec^{2}$, and $\mu_{0}^{c}$
is corrected for the inclination. $r_{e}$ and $r_{h}$ are respectively
the effective radius and disk scalelength derived  in arcsec, the corresponding
values expressed in kpc are derived by using $H_{0}=70 \mbox{km} \mbox{s}^{-1} 
\mbox{Mpc}^{-1}$. 
Third and sixth columns list the absolute errors derived by the fit routine.}
\begin{tabular}{ccccccc}
\tableline\tableline
P & Kn band & $\delta$ & kpc & I band & $\delta$ & kpc \\
\tableline
$\mu_{e}$ & 15.89 & 0.08 & & 18.86 & 0.05 & \\
$r_{e}$   & 1.41 & 0.04 & 0.3 & 0.71 & 0.02 & 0.1\\
$\mu_{0}$ & 15.12 & 0.04 & & 19.27 & 0.02 & \\
$\mu_{0}^{c}$ & 15.79 & 0.06 &  & 20.24 & 0.04 &\\
$r_{h}$ & 4.7 & 0.1 & 0.9 & 5.92 & 0.07 & 1.2  \\
$q_{b}$ & 0.99 & 0.01 & & 0.96 & 0.01 &  \\
$q_{d}$ & 0.54 & 0.01 & & 0.46 & 0.01 &\\
n & 0.57 & 0.11 & & 1.22 & 0.04 &  \\
$B/D$ & 0.122 & 0.005 &  & 0.107 & 0.003 & \\
\tableline
\end{tabular}
\end{center}
\end{table}

\newpage     

\end{document}